\documentclass[11pt]{article}
\usepackage{amsmath}
\usepackage{psfrag}
\usepackage{graphicx}

\def \mvspace {\vspace{-\abovedisplayskip}\vspace{-\belowdisplayskip}
\vspace{-\belowdisplayshortskip}\vspace{-\abovedisplayshortskip}}
\begin{document}
\pagestyle{empty}
\twocolumn[ \vspace*{-1.2cm}


\vspace*{36pt}
\center{
\parbox{16cm}{}{
\Large{\bf Multigrid solver for axisymmetrical 2D fluid
equations}\\
\vspace*{15pt}
 \large{\underline{Zoran Ristivojevic$^1$}, Zoran Lj.~Petrovi\' c$^1$}\\
~\\
 \small{\it $^1$Institute of Physics, University of Belgrade,
P.O. Box 68, 11080 Zemun, Serbia}} }\\
~\\
\vspace*{0pt}\center{
\parbox{14cm}{\small {We have developed an efficient algorithm for steady axisymmetrical
2D fluid equations. The algorithm employs multigrid method as well
as standard implicit discretization schemes for systems of partial
differential equations. Linearity of the multigrid method with
respect to the number of grid points allowed us to use $256\times
256$ grid, where we could achieve solutions in several minutes. Time
limitations due to nonlinearity of the system are partially avoided
by using multi level grids(the initial solution on $256\times 256$
grid was extrapolated steady solution from $128\times 128$ grid
which allowed using ``long'' integration time steps). The fluid
solver may be used as the basis for hybrid codes
for DC discharges.\\
\vspace*{1mm} }}}]

\noindent {\bf 1. Introduction}\\
Further understanding of basic processes in gas discharges and non-equilibrium
plasmas relies on comparisons of experimental results and predictions of
theoretical results which almost always have to be numerical calculations
including self consistent calculation of the spatial profile of electric field.
In all calculations it is critical to calculate the properties of electrons
which have large mobility and consequently ability to gain energy from the
electric field. Consequently electrons play a critical role in sustaining the
plasma by gas phase ionization. At the same time electrons have non-local or
non-hydrodynamic kinetics as their properties in the rapidly changing fields
may not be defined uniquely by the local electric field.  Thus, hybrid models
were developed to take into account high energy electrons originating from the
electrodes or created in very high electric fields close to the electrodes by
using a Monte Carlo technique while the bulk of electros at low energies is
accounted for by a fluid model.

As the need to describe accurately more and more complex geometries transition
from 1D to 2D ad 3D systems becomes increasingly demanding in terms of computer
time and complexity of equations that may lead to numerical problems such as
numerical diffusion and others.  In particular as the grid becomes denser the
computational demands increase as a square or cube of the number of grid
points. Thus, in order to model realistic structures and complex geometries
with proper relaxation of high energy electrons in space and time, one needs to
develop special numerical procedures to handle the complex task. In this paper
we describe one implementation of the multigrid technique which is far proved
to be the leading contender for optimal treatment of 3D systems with a large
number of grid points.

\noindent {\bf 2. Fluid equations and numerical algorithm}\\
Continuity equations for the electrons and the ions are [1]
\begin{align}
\label{kont_e}
\frac{\partial n_e}{\partial t}+\mathrm{\nabla\cdot}(n_e \mathbf v_e)=S_e,\\
\label{kont_p} \frac{\partial n_p}{\partial
t}+\mathrm{\nabla\cdot}(n_p \mathbf v_p)=S_p.
\end{align}
Momentum balance equations are
\begin{align}
\label{fluks_e}
\boldsymbol{\phi}_e=n_e \mathbf v_e=-n_e\mu_e\mathbf E-D_e\mathrm{\nabla}(n_e),\\
\label{fluks_p} \boldsymbol\phi_p=n_p \mathbf v_p=n_p\mu_p\mathbf
E-D_p\mathrm{\nabla}(n_p).
\end{align}

Electric potential is governed by the Poisson equation
\begin{equation}
\label{poisson} \mathrm{\nabla^2} V=-\frac{e}{\epsilon_0}(n_p-n_e),
\end{equation}
while the electric field is negative gradient of the potential:
\begin{equation}
\label{E} \mathbf E=-\mathrm{\nabla}V.
\end{equation}
In previous equations subscript index $e$ ($p$) refers to the
electrons (ions). Input parameters for the fluid equations are
source terms $S_e$ and $S_p$, mobilities $\mu_e(\mathbf E)$ and
$\mu_p(\mathbf E)$, diffusion coefficients $D_e(\mathbf E)$ and
$D_p(\mathbf E)$ and are supposed to be known( e.g. in hybrid models
they are provided from the lookup tables and Monte Carlo code).

We will solve a set of equations (\ref{kont_e})-(\ref{E}) for the
azimuthal symmetry ($f(r,z,\Theta)=f(r,z)$) and for a given set of
boundary conditions.

First we discretize system (\ref{kont_e})-(\ref{E}): we split the domain
$(r,z)\in [0,R]\times [0,d]$ into rectangles with equidistant radial grid
points $r_0,r_1,\ldots,r_{N_r}$ and non\-equ\-i\-dis\-tant axial grid points
$z_0,z_1,\ldots,z_{N_z}$. We also have midpoints between the grid points
$r_{i+1/2}$ and $z_{j+1/2}$. In the following we use indices $i,j,k$ for
$r$-coordinate, $z$-coordinate and for the time, respectively. In order to
allow long integration times the system must be discretized implicitly in time
[2]. Continuity equations (\ref{kont_e}) and (\ref{kont_p}) are discretized
``ba\-ckward in time''
\begin{eqnarray}
\label{kont2d_disk} \frac{n_{i,j}^{k+1}-n_{i,j}^k}{\Delta
t}+(\nabla_r\phi_r)_{i,j}^{k+1}+
(\nabla_z\phi_z)_{i,j}^{k+1}=S_{i,j}.
\end{eqnarray}

Momentum balance equations are of convection-diffusion type and it
is convenient to  discretize them by the Scharfetter--Gummel scheme
[2]:

\begin{align}
\label{sarfeter2dr_FG} \phi_{r,i+1/2,j}=\frac{D_{r,i+1/2,j}}{\Delta
r}\big(n_{i,j}G(\alpha_{i+1/2,j}^r)\\\notag-n_{i+1,j}F(\alpha_{i+1/2,j}^r)\big),\\
\label{sarfeter2dz_FG} \phi_{z,i,j+1/2}=\frac{D_{z,i,j+1/2}}{\Delta
z_j}\big(n_{i,j}G(\alpha_{i,j+1/2}^z)\\\notag-n_{i,j+1}G(\alpha_{i,j+1/2}^z)\big),
\end{align}
where
\begin{align}
\alpha_{i+1/2,j}^r=-s\frac{\mu_{r,i+1/2,j}}{D_{r,i+1/2,j}}(V_{i+1,j}-V_{i,j}),\\
\alpha_{i,j+1/2}^z=-s\frac{\mu_{z,i,j+1/2}}{D_{z,i,j+1/2}}(V_{i,j+1}-V_{i,j}).
\end{align}
(with $s=-1$ for the electrons and $s=1$ for the ions) and
\begin{align*}
&F(x)=\frac{x}{\exp\left[x\right]-1},
&G(x)=\frac{x\exp\left[x\right]}{\exp\left[x\right]-1}.
\end{align*}

Discretized form of the Poisson equation is
\begin{multline}
\label{poisson2ddiskretizovananormirana}
 \left(-2-\frac{(\Delta
r)^2(\Delta z_{j-1}+\Delta z_{j})}{\delta z_{j-1/2}\Delta z_{j-1}
\Delta z_{j}}\right)V_{i,j}^{k+1}\\+\frac{(\Delta r)^2}{\delta
z_{j-1/2}\Delta z_{j-1}}V_{i,j-1}^{k+1} +\frac{(\Delta r)^2}{\delta
z_{j-1/2}\Delta
z_{j}}V_{i,j+1}^{k+1}\\+\left(1-\frac{1}{2i}\right)V_{i-1,j}^{k+1}+
\left(1+\frac{1}{2i}\right)V_{i+1,j}^{k+1}\\-\frac{e}{\epsilon_0}n_0(\Delta
r)^2n_{e,i,j}^{\phantom{e,}k+1}+\frac{e}{\epsilon_0}n_0(\Delta
r)^2n_{p,i,j}^{\phantom{p,}k+1}=0,\\(i=1,2,\ldots,
N_r-1;j=1,2,\ldots,N_z-1),
\end{multline}
while along direction $r=0$ is
\begin{multline}
\label{poisson2ddiskretizovanaosanormirana} \left(-4-\frac{(\Delta
r)^2\left(\Delta z_{j-1}+\Delta z_{j}\right)}{\delta z_{j-1/2}\Delta
z_{j-1} \Delta z_{j}}\right)V_{0,j}^{k+1}+4V_{1,j}^{k+1}\\+
\frac{(\Delta r)^2}{\delta z_{j-1/2}\Delta
z_{j-1}}V_{0,j-1}^{k+1}+\frac{(\Delta r)^2}{\delta z_{j-1/2}\Delta
z_{j}}V_{0,j+1}^{k+1}\\-\frac{e}{\epsilon_0}n_0(\Delta r)^2
n_{e,0,j}^{\phantom{e,}k+1}+\frac{e}{\epsilon_0}n_0(\Delta r)^2n_{p,0,j}^{\phantom{p,}k+1}=0,\\
(j=1,2,\ldots,N_z-1).
\end{multline}

Discretized set of coupled equations connects concentrations and
potential at time $(k+1)\Delta t$ with values of concentrations and
potential at time $k\Delta t$. Since system of algebraic equations
(\ref{kont2d_disk})-(\ref{poisson2ddiskretizovanaosanormirana}) is
nonlinear we will solve it by applying the Newton--Raphson algorithm
[3]. Linearizing the system by introducing a vector $\mathbf
u=(n_e,n_p,V)^T$ and a correction $\delta\mathbf u$ of the same
vector (during Newton-Raphson iterations), we obtain a set of linear
equations for the correction vector:
\begin{align}
\label{nrjednacina} \hat a_{i,j}\delta\mathbf u_{i-1,j}+\hat
b_{i,j}\delta\mathbf u_{i,j}+ \hat c_{i,j}\delta\mathbf
u_{i+1,j}\\\notag+\hat d_{i,j}\delta\mathbf u_{i,j-1}+ \hat
e_{i,j}\delta\mathbf u_{i,j+1}=\mathbf f_{i,j},
\end{align}
\mvspace
\begin{multline}
\shoveleft{(i=0,1,\ldots N_r-1; j=1,2,\ldots N_z-1)\notag.}
\end{multline}
where $\hat a$ denotes that $a$ is a $3\times3$ matrix. Components
of the matrices from the previous equation are long expressions and
can be obtained straightforwardly. Equation (\ref{nrjednacina}) is
computationally very demanding and one has to solve it very
efficiently in order to get solutions of the fluid equations in a
reasonable time. Having in mind that fluid solver is usually
combined with Monte Carlo simulation in hybrid models, and that
after every cycle of Monte Carlo simulation one solves the set of
fluid equations, the importance of the efficiency is evident.

From known values of $n_e,n_p,V$ at $k\Delta t$ we calculate the
values at $(k+1)\Delta t$ by solving (\ref{nrjednacina}). After that
we iterate
(\ref{kont2d_disk})-(\ref{poisson2ddiskretizovanaosanormirana})
through time until the steady concentrations are obtained.

Since neither standard algorithms (e.g. Gauss elimination, LU
decomposition) nor iterative algorithms (Gauss-Seidel, Jacobi,
SOR)[3] are efficient enough, we have applied the multigrid method
[4] to solve (\ref{nrjednacina}). The latter method takes
$\mathcal{O}(N)$ operation for a linear system with $N$ unknowns,
while the former take at least $\mathcal{O}(N^2)$. Typically
$80\times 80$ or even coarser grids were used so far while we can
use much finer grids.

Moreover we have developed an improvement of the standard multigrid
algorithm. The main  idea consist of solving system
(\ref{nrjednacina}) on smaller discretization grid $N_r\times N_z$,
and then for the grid $2N_r\times 2N_z$ as an initial solution to
take extrapolated stationary solution from $N_r\times N_z$ and long
integration time step. In such way we go to finer and finer grids,
reaching the wanted number of grid points in the end (the finest one).

The solution on $N_r\times N_z$ was obtained taking arbitrary
initial $n_e,n_p$ and $V$ and the initial time step was usually
$\Delta t_1=1ns$ (or any time for which the system
(\ref{nrjednacina}) is convergent). After each time integration step
$\Delta t_k$, new time integration step is $\Delta t_{k+1}=n\Delta
t_k$ (usually $n=5$). It turns out that we can take greater and
greater integration time steps as we approach the stationary
solution. However if $t_{k+1}$ is too large (the norm of a
correction vector in Newton-Raphson iteration exceeds some upper
limit and the system is not robust for that time step), we go back
and choose $\Delta t_{k+1}=\Delta t_k$. In such way we approach the
stationary solution very fast. We call our method multilevel prolongation method.\\

\noindent {\bf 3. Results}\\
On the basis of glow discharge measurements in argon in cylindrical geometry
with $I=920\;\mu A$, $V=255.9\;V$; $d=1.1\;cm$, $R=2.7cm$, $pd=75\;Pa\cdot cm$
($p=0.514\;torr$)[6], we have have obtained an analytic form for the source
terms
$S_p(r,z)=0.35*10^{22}*\frac{\left(\frac{z}{L}\right)^2\exp\left(-50\left(\frac{z}{L}\right)^8\right)}
{1+\exp\left(a\frac{r}{R}-b\right)}\;m^{-3}s^{-1},
S_e(r,z)=1.75*10^{22}*\frac{\left(\frac{z}{L}-0.5\right)\exp\left(-40\left(\frac{z}{L}-0.5\right)^2\right)
\theta\left(\frac{z}{L}-0.5\right)}
{1+\exp\left(a\frac{r}{R}-b\right)}\;m^{-3}s^{-1}$.

Changing the parameters $a$ and $b$ one may obtain different radial
profiles which correspond to the cases of constricted and
non--constricted discharge. Other input parameters are
$\mu_e=\frac{30}{p}\; m^2 V^{-1}s^{-1}$ where $p$ is the pressure
given in $torr$; $D_e=\mu_e\frac{kT_e}{e}$ with $kT_e=1\;eV$;
$\mu_p$ was taken from [7]; $D_e=\mu_p\frac{kT_p}{e}$ with
$kT_p=0.026\;eV$. The boundary conditions are $V(r,0)=0$;
$V(r,d)=255.9\;V$; $n_e(r,0)=n_e(r,d)=0$; $n_p(r,0)=n_p(r,d)=0$;
$V(R,l)$ along the walls is interpolated linearly; $n_e=n_p=0$ along
the walls.

We have carried out calculations on a personal computer with processor Athlon
$3200+$ and $512$ Mb of RAM memory.

Number of Newton-Raphson iterations during the solving of nonlinear
system was less than $10$. Newton-Raphson iterations are stopped
when the sum of the corrections was less than $10^{-6}$. Time
evolution are stopped when the relative change in densities become
less than $10^{-6}$. The overall integration time for the stationary
solution was of the order of $1s$ which is unimportant for our
algorithm which advances through time by a factor $n$.

\begin{figure}[t]
\centering
\includegraphics[width=0.8\linewidth]{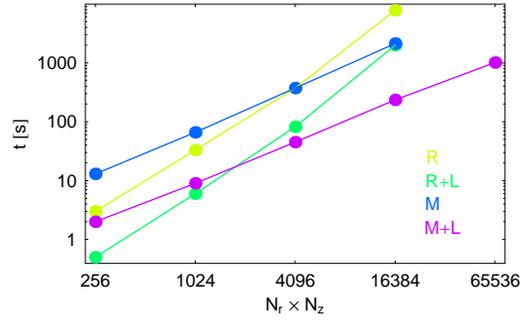}
\caption{Execution times in seconds for different integration
strategies. $R$ denotes results obtained by iterative algorithm, $M$
by multigrid, while $L$ denotes multilevel prolongation method.
Grids have $N_r=N_z.$}\label{slikapot}
\end{figure}

Execution times are given in Fig.\ref{slikapot}. We can easily see that the
slope of the multigrid solvers (without (M) and with our acceleration (M+L)) is
different that for the iterative method which confirms different efficiencies
of these methods.

%

\begin{figure}[t]
\centering
\includegraphics[width=0.99\linewidth]{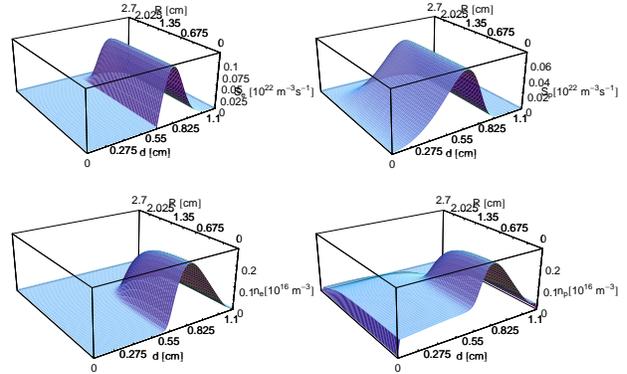}
\caption{Source terms and densities for $a=30$, $b=24$ for grid size
$64\times 64$.}\label{slikanenp1}
\end{figure}

\begin{figure}[t]
\centering
\includegraphics[width=0.99\linewidth]{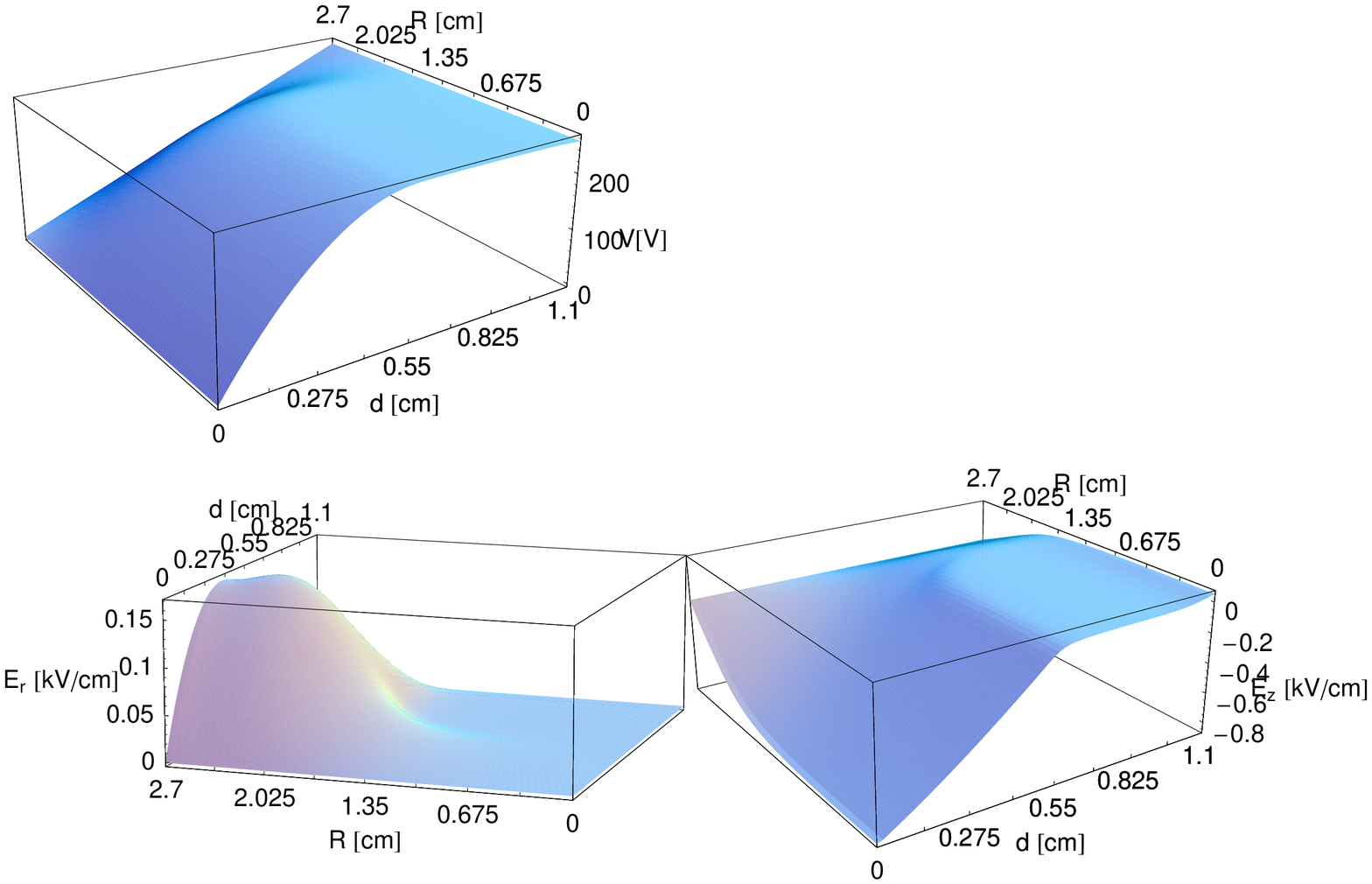}
\caption{Potential and electric field for $a=30$, $b=24$ for grid
size $64\times 64$.}\label{slikaVE1}
\end{figure}

\begin{figure}[t]
\centering
\includegraphics[width=0.99\linewidth]{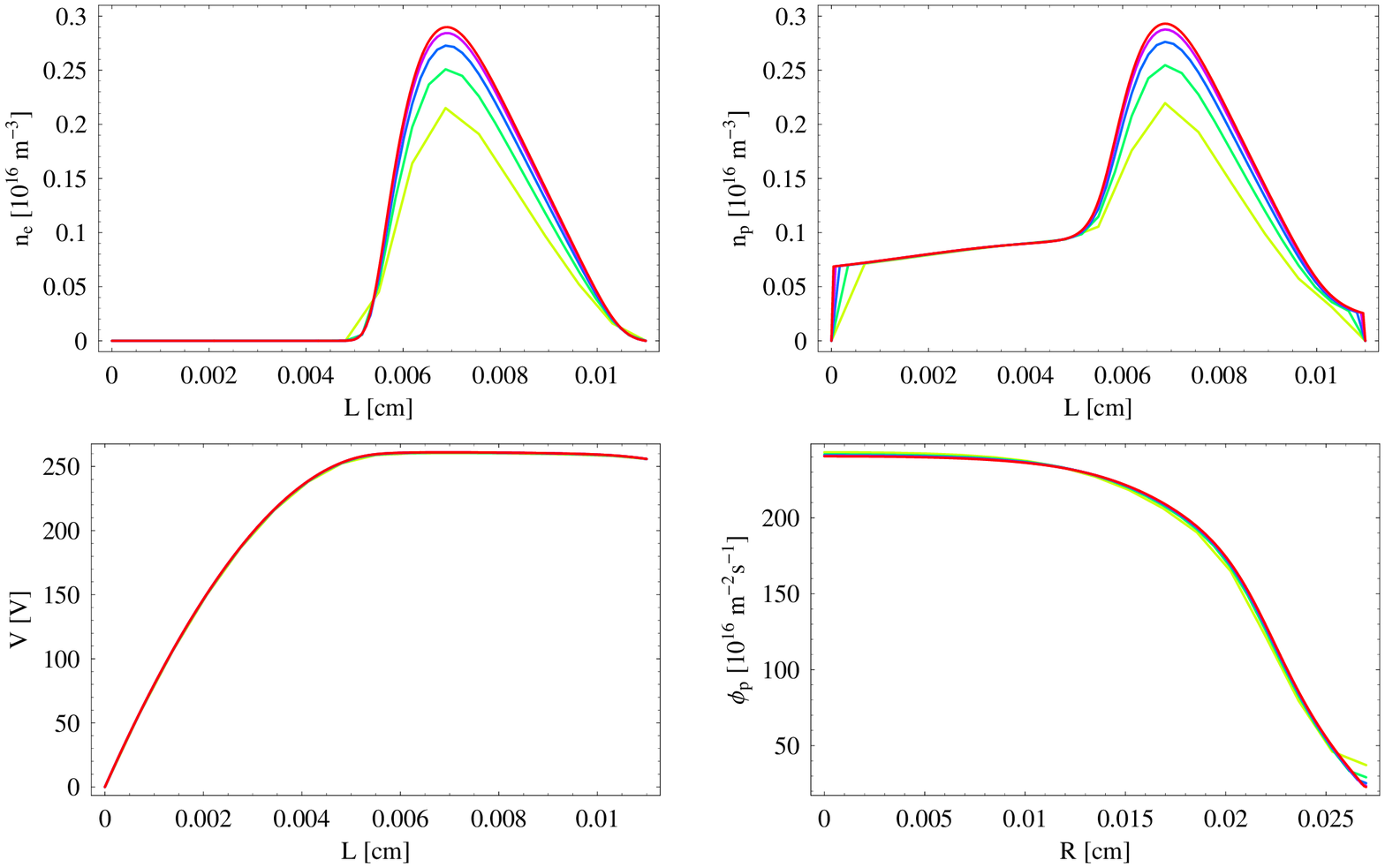}
\caption{Potential and electric field for $a=30$
$b=24$.}\label{slika1}
\end{figure}

\begin{figure}[t]
\centering
\includegraphics[width=0.99\linewidth]{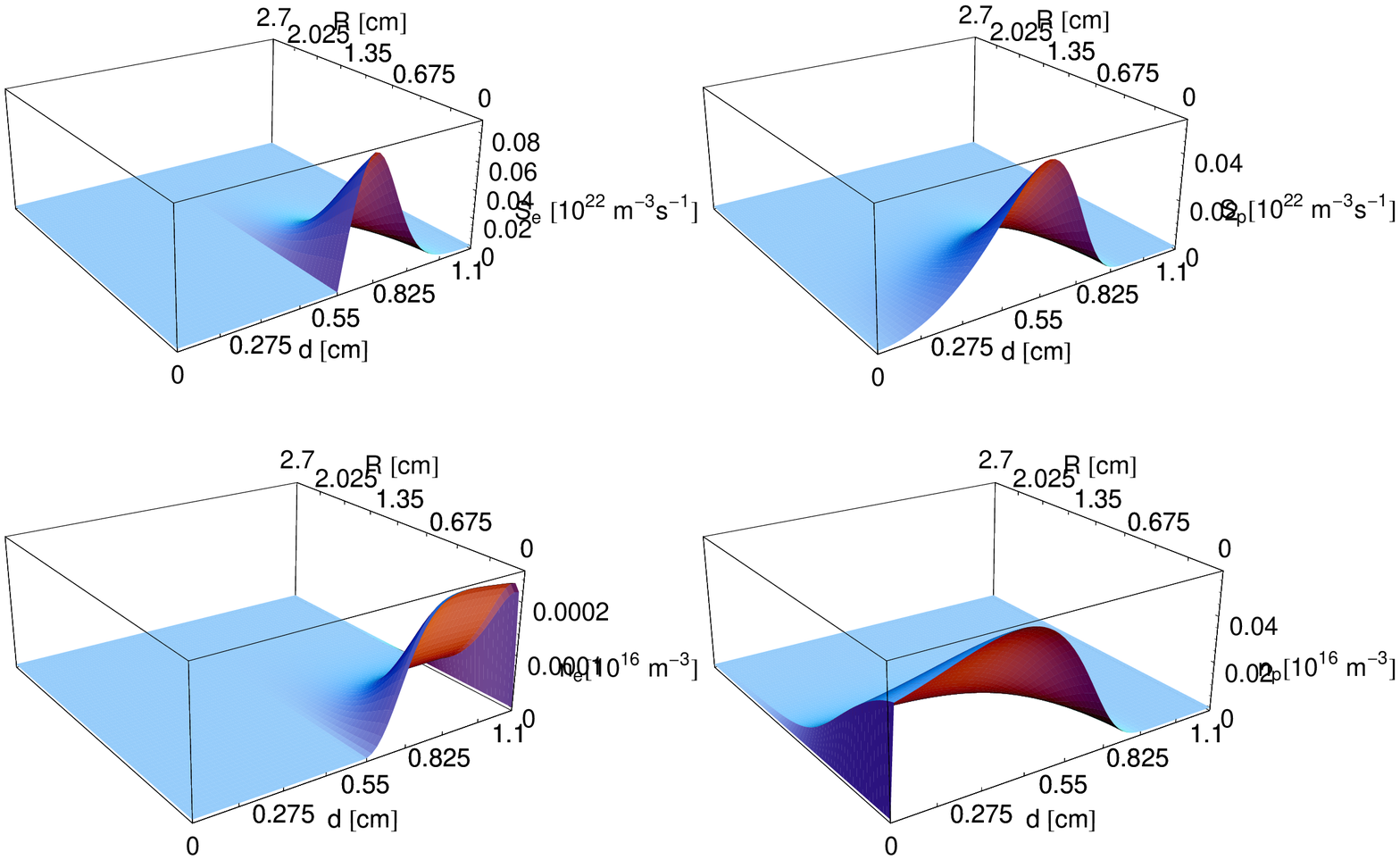}
\caption{Source terms and densities for $a=10$, $b=1$ for grid size
$64\times 64$.}\label{slikanenp2}
\end{figure}

\begin{figure}[t]
\centering
\includegraphics[width=0.99\linewidth]{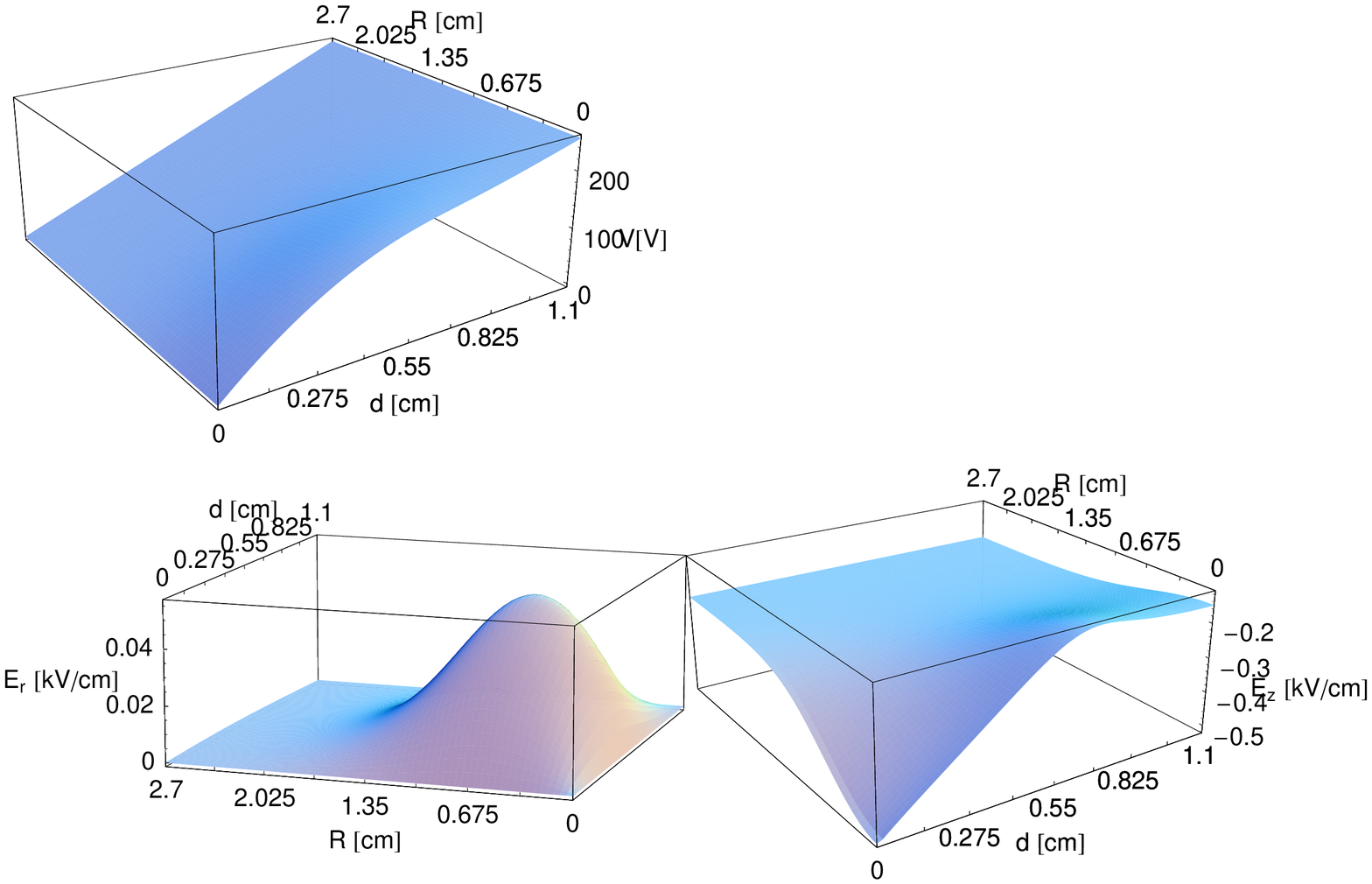}
\caption{Potential and electric field for $a=10$, $b=1$ for grid
size $64\times 64$.}\label{slikaVE2}
\end{figure}

\begin{figure}[t]
\centering
\includegraphics[width=0.99\linewidth]{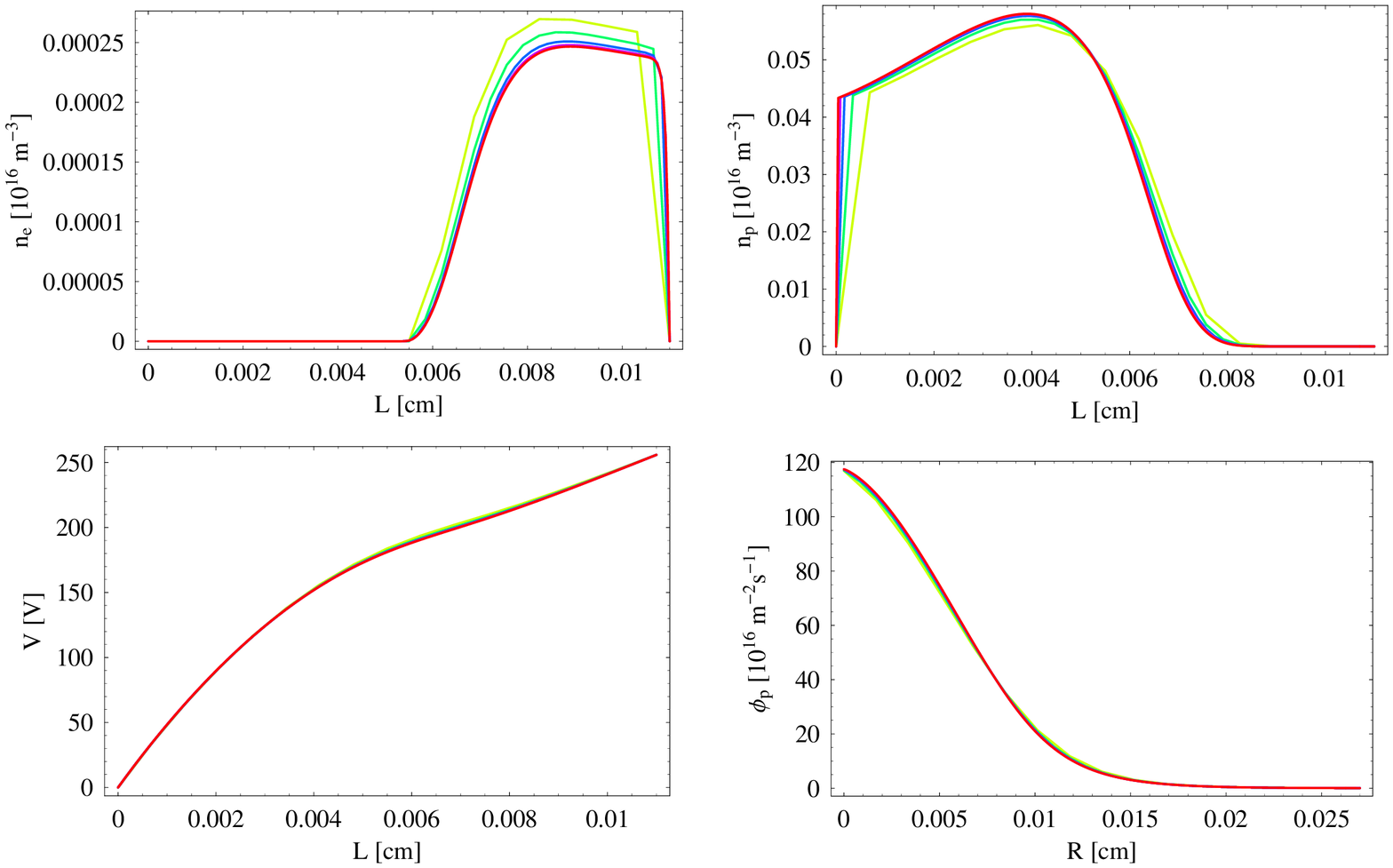}
\caption{Potential and electric field for $a=10$
$b=1$.}\label{slika2}
\end{figure}

Fig.\ref{slikanenp1}--Fig.\ref{slika1} and
Fig.\ref{slikanenp2}--Fig.\ref{slika2} show the solution for $a=30,
b=24$ and $a=10, b=1$, respectively. We can see in
Fig.\ref{slika1},Fig.\ref{slika2} that the electron and ion
densities are highly sensitive to the grid size. In the first case
both densities increase with the grid, while in the second case the
electron density decreases. The ion current to the cathode in the
first case is $I=0.59 mA$ and in the second case $I=0.035 mA$ and
are only very weakly dependant of the grid size.\\

\noindent {\bf  4. Conclusions}\\
We have developed an efficient method for 2D fluid equations in cylindrical
geometry. We have applied the multigrid to our knowledge for the first time to
solve a system of equations for an obstructed dc glow discharge and with
implementation for multi level prolongation approach. Tremendous acceleration
in respect to the iterative methods is obtained by applying multigrid method
combined with multilevel prolongation method. From the solution we may conclude
importance of the grid size: is not so important for the ion flux to the
cathode, but for densities it is important.

A question which arises is how fine grid may be imposed by the
physics of the problem. The answer should be probably not finer than the mean free
path for the ionization which may vary with the external parameters
of the discharge.
\\

\noindent {\bf 5. References}

[1] J. D. P. Passchier, Numerical Fluid Models for RF discharges,
PhD Thesis, Utrecht, 1994.

[2] A. Fiala, L. C. Pitchford, J. P. Boeuf, Phys. Rev. E \textbf{49}
(1994) 5607.

[3] W. H. Press, S. A. Teukolsky, W. T. Vetterling, B. P. Flannery,
\emph{Numerical Recipes in C}, Second Edition (Cambridge University
Press, 1992).

[4] U. Trottenberg, C. W. Oosterlee, A. Sch\"{u}ller,
\emph{Multigrid}, (academic press, 2001).

[5] Z. Donk\' o, private communication.

[6] D. Mari\' c et all, J. Phys. D: Appl. Phys. \textbf{36} (2003)
2639.

[7] A. V. Phelps, Z. Lj. Petrovi\'c, Plasma Sources Sci. Technol.
\textbf{8} (1999) R21.

\end{document}